\documentclass[a4page,10pt]{article}
\usepackage{amssymb,amsmath}
\numberwithin{equation}{section}

\begin{document}

\title{ A Simple Quantum-Mechanical Model of Spacetime I: Microscopic Properties of Spacetime}

\author{Jarmo M\"akel\"a\footnote{Vaasa University of Applied Sciences,
Wolffintie 30, 65200 Vaasa, Finland (email: jarmo.makela@puv.fi)}} 

\maketitle

\begin{abstract}

This is the first part in a series of two papers, where we consider a specific microscopic 
model of spacetime. In our model Planck size quantum black holes are taken to be the 
fundamental building blocks of space and time. Spacetime is assumed to be a graph, where 
black holes lie on the vertices. In this first paper we construct our model in details, 
and show how classical spacetime emerges at the long distance limit from our model. We 
also consider the statistics of spacetime. 
 
\end{abstract}



\section{Introduction}

      Gravitation is the most interesting of all known fundamental interactions of nature.
Its main interest lies in the fact that the properties of gravitation are closely linked 
to the properties of space and time. Indeed, we have learned from Einstein's general theory 
of relativity that gravitation is interaction between matter fields and the geometry of 
spacetime. Because of the intimate relationship between gravitation, space and time it is
unavoidable that any plausible theory of gravity always involves a theory of space and time.

    This paper is the first part in a series of papers, where we consider a proposal
 for a simple model of spacetime, which takes into account the quantum effects on its structure 
and properties. In other words, we consider a quantum mechanical model of spacetime. It is 
generally believed that quantum effects of gravity will dominate at length scales which are
 characterized by the so called Planck length
\begin{equation}
\ell_{Pl} := \sqrt{\frac{\hbar G}{c^3}} \approx 1.6\times 10^{-35}m.
\end{equation}
By means of our model we consider the structure and the properties of spacetime at the 
Planck length scale, and how classical spacetime as such as we know it from Einstein's 
general relativity emerges out from quantum spacetime at macroscopic length scales. An 
advantage of our model is that it reproduces, in a fairly simple manner, the "hard facts" 
of gravitational physics as such as we know them today. For instance, at macroscopic length 
scales our model reproduces Einstein's field equation with a vanishing cosmological 
constant. In the semiclassical limit our model reproduces, for low temperatures, the 
Hawking and the Unruh effects. Our model also makes some novel predictions, which
have already been anticipated, on various grounds, by other authors. Among other things
our model predicts that area has a discrete spectrum with an equal spacing, and that in
high enough temperatures black hole entropy depends logarithmically on the event
horizon area, instead of being proportional to the area.

    The basic idea of our model may be traced back to Jacobson's very important discovery
of the year 1995 that Einstein's field equation may actually be understood as a sort of
thermodynamical equation of state of spacetime and matter fields \cite{yksi, kaksi}. More precisely, Jacobson 
considered the flow of heat through the local past Rindler horizon of an accelerated 
observer, and he {\it assumed} that when heat flows through a finite part of the horizon, 
then the amount of entopy carried through that part is, in natural units, one-quarter of 
the decrease in the area of that part. Identifying the Unruh temperature of the 
accelerated observer as the temperature assigned to the heat flow Jacobson found, by 
using the Raychaudhuri equation, that Einstein's field equation is a straightforward 
consequence of the fundamental thermodynamical relation
\begin{equation}
\delta Q = T\,dS
\end{equation}
between the heat $\delta Q$, absolute temperature $T$, and entropy $dS$ carried through 
the horizon. So it appears that Eq.(1.2), which was first introduced by Clausius in 1865
as a thermodynamical definition of the concept of entropy \cite{kolme}, is not only one of 
the most important equations of thermodynamics, but it is also the fundamental equation of
gravitation. When horizon area is interpreted as a measure of entropy associated with 
the horizon, classical general relativity with all of its predictions is just one of the
consequences of Eq.(1.2). The close relationship between gravitation and thermodynamics was 
already suggested by the Bekenstein-Hawking entropy law which states that black hole has 
entropy $S_{BH}$ which, in natural units, is one-quarter of its horizon area or, in SI 
units \cite{nelja, viisi},
\begin{equation}
S_{BH} = \frac{1}{4}\frac{k_Bc^3}{\hbar G}A,
\end{equation}
but Jacobson's discovery makes the case clear: On the fundamental level classical gravity 
{\it is} thermodynamics of spacetime and matter fields \cite{kuusi}.

    If the thermodynamical interpretation of gravity turns out to be correct, the problem of 
quantum gravity, or a theory which would bring together quantum mechanics and general relativity,
becomes obvious: Instead of attempting to quantize Einstein's classical general relativity as 
if it were an ordinary field theory, we should try to identify the fundamental, microscopic
constituents of spacetime, and to postulate their quantum mechanical and statistical 
properties. These fundamental constituents play in the fabric of spacetime a role similar to that of 
atoms in the fabric of matter. Presumably those constituents are Planck size objects. 
The thermodynamical properties of spacetime follow from the statistics of its constituents,
and at appropriate length scales those thermodynamical properties should reproduce, at 
least in low temperatures, the well known effects of classical and semiclassical gravity. 
Although such thoughts were already expressed by Misner, Thorne and Wheeler in the inspired 
final chapter of their book \cite{seitseman}, the approach described above has never, to 
the best knowledge of the author, been attempted really systematically. Loop quantum gravity bears
some resemblance to such an approach, but its roots lie in the attempts to quantize 
Einstein's general relativity as if it were an ordinary field theory, and its statistical 
and thermodynamical aspects are still unknown \cite{kahdeksan}. So does the approach based 
on the so called causal triangulations of spacetime, but that approach, in turn, takes its ideas 
from the attempts to quantize gravity by means of the path integral methods applied for Regge 
calculus \cite{yhdeksan}. On the attempts to approach quantization of gravity by means of postulating some 
fundamental properties for spacetime one should mention, among other things, Sorkin's causal 
set theory \cite{kymmenen} and, in particular, the so called Barret-Crane model \cite{yksitoista}.
There are also some other approaches of this type \cite{kaksitoista}.

     The first problem is to identify the fundamental constituents of spacetime. At the 
moment we have no idea what these fundamental constituents might actually be, but some 
hints may be gained by means of some general, quantum mechanical arguments. For instance,
suppose that we have closed a particle inside a box whose edge length is one Planck length
$\ell_{Pl}$. In that case it follows from Heisenberg's uncertainty principle that the 
momentum of the particle has an uncertainty $\Delta p\sim \hbar/\ell_{Pl}$. In the 
ultrarelativistic limit this uncertainty in the momentum corresponds to the uncertainty 
$\Delta E\sim c\Delta p$ in the energy of the particle. In other words, inside a box with
an edge length equal to $\ell_{Pl}$ we have closed a particle whose uncertainty in its 
energy is about the same as the so called Planck energy
\begin{equation}
E_{Pl} := \sqrt{\frac{\hbar c^5}{G}}\approx 2.0\times 10^9 J.
\end{equation}
This amount of energy, however, is enough to shrink the spacetime region surrounding the 
cube into a {\it black hole} with a Schwarzschild radius equal to around one Planck length.
So it is possible that one encounters with Planck size black holes when probing spacetime at
the Planck length scale. Such an idea is far from new. For instance, Misner, Thorne and 
Wheeler write in their book, how "[gravitational] collapse at the Planck scale of distance
is taking place everywhere and all the time in the quantum fluctuations in the geometry and,
one believes, the topology of spacetime."\cite{seitseman} These kind of sentences immediately bring into 
one's mind mental images of wormholes and tiny black holes furiously bubbling as a sort of
spacetime foam. Unfortunately, the idea of spacetime being made of Planck size black holes 
has never been taken very far \cite{kolmetoista, neljatoista}.

        In our model we take most seriously the idea that spacetime might really be made 
of Planck size black holes. In other words, we consider Planck size quantum black holes 
as the fundamental constituents of space and time. We postulate certain properties for 
such black holes, and from these postulates we deduce the macroscopic properties of space
and time. Although our model is really very simple and straightforward, it involves lots 
of new ideas and concepts, and careful penetration is therefore required from the reader.  
To fully grasp the physical content of our model the both papers in our series should be
read slowly and carefully, word by word, from the beginning to the end.

     We begin the construction of our model in Section 2 of this paper by defining the concept of a 
microscopic quantum black hole. The fundamental, undefined quantity associated with a 
microscopic quantum black hole is its {\it horizon area}. Following a proposal already 
made by Bekenstein in 1974 \cite{viisitoista} we shall assume that the spectrum of the horizon area operator
is discrete with an equal spacing. Horizon area is the only observable associated with a 
microscopic black hole, and all properties of space and time
may ultimately be reduced back to the horizon area eigenvalues of the Planck size black 
holes constituting spacetime. Other possible quantities associated with black holes,
such as the mass, for instance, have no relevance at the Planck scale of distance. Actually,
even the concept of distance is abandoned. The fundamental concept is horizon area, and the
concept of distance, as well as the other metric concepts associated with spacetime, arise 
as sort of statistical concepts at macroscopic length scales. 

  In Section 3 we construct a precise definition of spacetime. Our definition is based on 
the concept of {\it graph}, which is one of the simplest concepts of mathematics. In very 
broad terms, we postulate that spacetime is a certain kind of graph, where black
holes lie on the vertices. When constructing the mathematically precise definition of the graph which
ties together the black holes lying on its vertices, we are forced to introduce several new
concepts, which are graph theoretic analogues of the corresponding concepts of classical spacetime.
As the last step in the construction of the graph in question we introduce the concept of {\it proper
spacetime lattice}, which may be described as a sort of graph theoretic analogue of a differentiable
manifold. Our definition of spacetime is constructed in such a way that 
it makes possible to define the concept of {\it two-dimensional subgraph}, which is an analogue 
of the concept of two-surface in classical spacetime. 
   
     In Section 4 we proceed to the statistics of spacetime. At this stage two-dimesional 
subgraphs take a preferred role. We express five {\it independence}- and two 
{\it statistical} 
postulates for the black holes lying on the vertices of a two-dimensional subgraph. Among other
things our postulates state, in very broad terms, that the area of a spacelike two-graph
is proportional to the sum of the horizon areas of the black holes lying on its vertices,
and that the macroscopic state of a two-dimensional subgraph is determined by its area, whereas
its microscopic states are determined by the combinations of the horizon area eigenstates
of the holes. In other words, area takes the role of energy as the fundamental statistical 
quantity in our model in a sense that the microscopic and the macroscopic states of a 
two-dimesional subgraph are labelled by the horizon areas, instead of the energies, of its 
constituents. Our postulates imply, among other things, that in the so called low temperature limit, where 
most black holes are on their ground states, the natural logarithm of the number of 
microstates corresponding to the same macroscopic state of the two-dimensional subgraph is 
directly proportional to the area of that subgraph. This is a very important conclusion, and
it is closely related to the Bekenstein-Hawking entropy law.

      In Section 5 we consider how classical spacetime as such as we know it from Einstein's
general theory of relativity emerges from quantum spacetime at macroscopic length 
scales. When worked out with a mathematical precision, the process of reconstruction of 
classical out of quantum spacetime is a rather long one. However, the basic idea may be 
summarized, again in very broad terms, by saying that at macroscopic length scales for every
vertex $v$ of spacetime there exists a subgraph $G_v$ in such a way that $G_v$ may be 
approximated, in a certain very specific sense, by a geometrical four-simplex $\sigma_v$ 
such that the areas of the two-faces, or triangles, of the four-simplex $\sigma_v$ are 
determined by the horizon areas of the black holes in $G_v$. In our model the geometrical 
four-simplex $\sigma_v$ plays the role analogous to that of tangent space in classical
general relativity. Every four-simplex $\sigma_v$ is equipped with a flat Minkowski metric.
It is a specific property of four-simplices that each four-simplex has an equal number of 
triangles and edges. Because of that the lengths of edges may be expressed in terms of the 
areas of triangles which, moreover, are determined by the horizon areas of the Planck size 
black holes. In other words, the metric properties of spacetime may be reduced to the 
horizon area eigenstates of its quantum constituents. However, it should be noted that the 
metric of spacetime is a statistical quantity, which is meaningful at macroscopic length 
scales only. After introducing the concept of metric in spacetime, it is possible to define other
concepts familiar from classical general relativity, such as the Christoffel symbols, the 
Riemann and the Ricci tensors, and so forth.

   We close our dicussion with some concluding remarks in Section 6.

\maketitle

\section{Microscopic Quantum Black Holes}

   Black holes are extremely simple objects: After a black hole has settled down, it has 
just three classical degrees of freedom, which may be taken to be the mass, the angular momentum, 
and the electric charge of the hole \cite{kuusitoista}. Because of their simplicity, microscopic black holes are 
ideal candidates for the fundamental constituents, or atoms, of spacetime. The values taken
by just three quantities are enough to specify all of the properties of an individiual 
microscopic black hole, and in terms of these properties one may attempt to explain all of
the properties of spacetime.

   It is obvious that microscopic black holes should obey the rules of quantum mechanics.
When constructing a quantum mechanical model of spacetime out of microscopic black holes, 
the first task is therefore to find the spectra of the three classical degrees of freedom,
or obsevables, of the hole \cite{seitsemantoista}. In what follows, we shall simplify the problem further, and we 
shall assume that each microscopic black hole acting as an atom of spacetime has just {\it
one} classical degree of freedom, which may be taken to be the mass of the hole. In other 
words, we shall assume that spacetime is made of microscopic, non-rotating, electrically 
neutral black holes. This means that in our model spacetime is made of microscopic 
Schwarzschild black holes.

      There have been numerous attempts to quantize the mass of the Schwarzschild black 
hole. Quite a few of them have reproduced a result, which is known as {\it Bekenstein's
proposal}. According to this proposal, which was expressed by Bekenstein already in 1974,
the event horizon area of a black hole has an equal spacing in its spectrum 
\cite{viisitoista}. More 
precisely, Bekenstein's proposal states that the possible eigenvalues  of the event 
horizon area $A$ of a black hole are of the form:
\begin{equation}
A_n = n\gamma\ell_{Pl}^2,
\end{equation}
where $n$ is a non-negative integer, $\gamma$ is a numerical constant of order one, and 
$\ell_{Pl}$ is the Planck length of Eq.(1.1) \cite{kahdeksantoista}.

       There are indeed very good reasons to believe in Bekenstein's proposal. First of all,
we have dimensional arguments: When a system is in a bound state, the eigenvalues of any 
quantity tend to be quantized in such a way that when we write the natural unit for the 
quantity under consideration in terms of the natural constants relevant for the system, 
then (at least in the semiclassical limit), $\hbar$ is multiplied by an integer in the 
spectrum. For Schwarzschild black holes the only relevant natural constants are $\hbar$, 
$G$ and $c$, and we find that $\ell^2_{Pl}$ is the natural unit of horizon area. Since 
$\ell^2_{Pl}$ is proportional to $\hbar$, one expects that $\ell_{Pl}^2$ is multiplied by
an integer in the spectrum. In other words, one expects that the horizon area spectrum is 
of the form given in Eq.(2.1).

      Another argument often used to support Bekenstein's proposal is Bekenstein's 
observation that horizon area is an {\it adiabatic invariant} of a black hole \cite{viisitoista}.
 Loosely speaking, this means that the event horizon area remains invariant under very slow 
external perturbations of the hole. Adiabatic invariants of a system, in turn, are 
given by the {\it action variables} $J$ of the system. At least in the semiclassical limit
the eigenvalues of the action variables of any system are of the form:
\begin{equation}
J_n = 2\pi n\hbar,
\end{equation}
where $n$ is an integer. Because the event horizon area is an adiabatic invariant of a black
hole, it is possible that the event horizon area is proportional to one of the action 
variables of the hole. If this is indeed the case, one arrives, again, at the spectrum 
given by Eq.(2.1).

    In this paper we take Bekenstein's proposal most seriously. We shall {\it assume} that
the possible eigenvalues of the event horizon areas of the microscopic Schwarzschild 
black holes constituting spacetime are of the form:
\begin{equation}
A_n = (n + \frac{1}{2})32\pi\ell^2_{Pl},
\end{equation}
where $n=0,1,2,...$. This kind of a spectrum is not merely 
idle speculation, but it is really possible to construct, using the standard rules of 
quantum mechanics, a plausible quantum mechanical model of the Schwarzschild black hole, 
which produces the area spectrum of Eq.(2.3). \cite{yhdeksantoista} At this point we shall not need the details 
of that model, but the horizon area spectrum of Eq.(2.3) is sufficient for our 
considerations.

    Consider now Eq.(2.3) in more details. First of all, it is instructive to compare 
Eq.(2.3) with the energy spectrum of the field quanta in ordinary quantum field theories.
In ordinary quantum field theories in flat spacetime the standard rules of quantum 
mechanics imply the result that matter field may be thought to be constructed of 
particles, or field quanta, such that the possible energy eigenvalues of particles with
momentum $\vec k$ are of the form:
\begin{equation}
E_{n,\vec k} = (n + \frac{1}{2})\hbar\omega_{\vec k},
\end{equation}
where $n=0,1,2,...$, and $\omega_{\vec k}$ is the angular frequency corresponding to the 
momentum $\vec k$. As one may observe, there is an interesting similarity between the 
Eqs.(2.3) and(2.4): The event horizon area is quantized in exactly the same way as is the 
energy of the particles of the field. In other words, event horizon area takes the place of 
energy in our model, and we have just replaced $\hbar\omega_{\vec k}$ by $32\pi\ell_{Pl}^2$.
Indeed, one is strongly tempted to speculate that, in the fundamental level, the "quanta" 
of the gravitational field are actually microscopic quantum black holes, and that in quantum
gravity area plays a role similar to that of energy in ordinary quantum field theories.

     Since the Schwarzschild mass $M$ is the only degree of freedom of the classical 
Schwarzschild black hole, one may be interested in the mass spectrum of the microscopic 
quantum black holes as well. Because the horizon area $A$ of the Schwarzschild black hole 
is related to its mass such that:
\begin{equation}
A = \frac{16\pi G^2}{c^4}M^2,
\end{equation}
it follows from Eq.(2.3) that the eigenvalues of $M$ are of the form:
\begin{equation}
M_n = \sqrt{2n + 1}\,M_{Pl},
\end{equation}
where
\begin{equation}
M_{Pl} := \sqrt{\frac{\hbar c}{G}}\approx 2.2\times 10^{-8} kg
\end{equation}
is the Planck mass. However, as we shall see in Section 4, the Schwarzschild mass of a 
microscopic quantum black hole does not, when the hole is considered as a constituent of 
spacetime, have any sensible physical interpretation.

      In this paper we shall assume that microscopic black holes obey the standard rules of 
quantum mechanics. One of the consequences of this assumption is that the physical states 
of an individual black hole constitute a Hilbert space, which we shall denote by 
$\cal{H}_{BH}$. In this Hilbert space operates a self-adjoint event horizon area operator 
$\hat A$ with a spectrum given by Eq.(2.3), and $\cal{H}_{BH}$ is spanned by the normalized
eigenvectors $\vert \psi_n\rangle$ ($n=0,1,2,..$) of $\hat A$. In other words, an arbitrary
element $\vert \psi\rangle$ of $\cal{H}_{BH}$ may be written in the form:
\begin{equation}
\vert \psi\rangle = \sum_{n=0}^\infty c_n\vert \psi_n\rangle,
\end{equation}
where $c_n$'s are complex numbers. 

\maketitle

\section{A Model of Spacetime}

\subsection{The Concept of Graph}

     Our next task is to construct a mathematically precise model of spacetime made of black
holes. This is a very challenging task, because when constructing such a model we must 
think carefully, what kind of properties do we expect spacetime to possess on the fundamental
level. For instance, it is possible that at the Planck length scale spacetime does not 
have any metric properties, but the metric properties arise as statistical properties of 
spacetime at length scales very much larger than the Planck length scale. Hence, it appears 
that we must abandon, at the Planck length scale, the metric structure, and also some other 
structures familiar from classical general relativity. In our model we shall abandon the 
differentiable structure, and even the manifold structure.

   If we abandon even the manifold structure, then what are we left with? Just a set of 
points. To abandon even the last shreds of hopes of being able to construct a manifold 
structure for spacetime we shall assume that this set of points is countable. No 
manifold structure may be constructed for a countable set, because no countable set is
homeomorphic to $\Re^n$ for any $n$.

   If  we have just a countable set of points, then what can we do with those points? Well,
we may arrange them in pairs, for instance. This idea brings us to {\it graph theory} \cite{kaksikymmenta}. 
Mathematically, a {\it graph} $G$ is defined as an ordered triple $({\cal V}, E, f)$, where 
${\cal V}$ is the set of {\it vertices} of the graph, $E$ is the set of its {\it edges}, and
$f$ is a map from $E$ to the set of unordered pairs of ${\cal V}$. In other words, for each
edge there is an unordered pair of the elements of ${\cal V}$, and therefore an edge may be 
understood as an unordered pair $\lbrace u,v\rbrace$ of the vertices of the graph. An edge 
with vertices $u$ and $v$ is denoted by $uv$. The set ${\cal V}$ of the vertices of the 
specified graph $G$ is sometimes denoted by ${\cal V}(G)$, and the set $E$ of its edges by
$E(G)$. The sets ${\cal V}(G)$ and $E(G)$ are assumed to be finite. A {\it subgraph} $G'$ 
of the given graph $G$ is a graph whose vertex and edge sets ${\cal V}(G')$ and $E(G')$
are subsets of those of the graph $G$.

   Graph theory is an important branch of mathematics which, probably because of an extreme
simplicity of its concepts and ideas, has important applications even outside pure 
mathematics. The most important practical applications of graph theory lie in network 
analysis. Graph theory also has some unsolved problems of practical value, such as the so 
called Travelling Salesman problem.

\subsection{Elementary Spacetime Lattice}

\subsubsection{Labelling of the Vertices}

In this paper we shall denote the vertices of a graph acting as a model of spacetime by $v(k^0,k^1,k^2,k^3)$,
where the numbers $k^\mu$ ($\mu = 0, 1, 2, 3$) are integers. We shall usually employ a shorthand notation, where
we denote:
\begin{equation}
v(k^0,k^1,k^2,k^3) := v(k^\mu).
\end{equation}
In other words, each vertex of our graph is labelled by four integers $k^\mu$. This is an attempt to bring 
something resembling the manifold structure of classical spacetime into our model: In classical spacetime 
each point is identified by four reals $x^\mu$ ($\mu = 0, 1, 2, 3$), whereas in our model each vertex is 
identified by four integers $k^\mu$. Since the integers $k^\mu$ are analogous to the coordinates $x^\mu$ of
the points of classical spacetime, we shall call the integers $k^\mu$ as the {\it coordinates} of the vertices
of our graph.

  \subsubsection{Elementary Spacetime Lattice (ESL)}

   As the first step towards the definition of a mathematicall precise model of spacetime we define the concept
of {\it elementary spacetime lattice} (ESL) as a graph $G$ with the following properties:

  (i) There exists an integer $N\ge 2$ such that $v(k^\mu)\in \mathcal{V}(G)$ if and only if 
$k^\mu \in \lbrace 1, 2, 3,..., N\rbrace$ for all $\mu = 0, 1, 2, 3$.

  (ii) $v(k^\mu)v(k'^\nu)\in E(G)$ if and only if 
       \begin{equation}
       max\lbrace\vert k'^\mu - k^\mu\vert\rbrace = 1
      \end{equation}
 for all $\mu = 0, 1, 2, 3$.

   Our definition of ESL implies that all coordinates $k^\mu$ of the vertices $v(k^\mu)$ of ESL take all integer
values from 1 to $N$, and the vertices $v(k^\mu)$ and $v(k'^\mu)$ of an arbitrary edge $v(k^\mu)v(k'^\mu)$ of 
ESL are different. Moreover, the condition (ii) in our definition implies that $v(k^\mu)v(k'^\mu)$ is an edge of
ESL if and only if the coordinates $k^\mu$ and $k'^\mu$ of the vertices $v(k^\mu)$ and $v(k'^\mu)$ differ from
each other by exactly one (and not more) at least once.

  \subsubsection{An Example of ESL}

  The simplest possible example of an ESL is a graph with vertices $v(k^0,k^1,k^2,k^3)$ such that all coordinates
$k^\mu$ ($\mu = 0, 1, 2, 3$) take the values 1 and 2 only, and every vertex is connected to every other vertex by
an edge. Such an ESL contains $2^4 = 16$ vertices and $\left(\begin{array}{cc}16\\2\end{array}\right) = 120$ edges.
In our subsequent discussion we shall call the ESL in question as a {\it unit cell}.

  \subsubsection{Path}

  In graph theory, {\it walk} is defined as an alternating sequence of vertices and edges, in which each vertex is
incident to the two edges that precede and follow it in the sequence, and the vertices that precede and follow an 
edge are the vertices of that edge. A walk is a {\it path}, if its first and last vertices are different. The 
{\it length} of a path is the number of its edges.

  \subsubsection{Distance}

   According to a standard definition, the {\it distance} $d_G(u,v)$ between two vertices $u$ and $v$ in a 
graph $G$ is the length of the shortest path between them. The subscript $G$ is usually dropped, when there is
no danger of confusion. When $u$ and $v$ are identical, their distance is 0. When $u$ and $v$ are unreachable
from each other, their distance is defined to be infinity $\infty$.

   We shall prove later the following theorem:

   {\bf Theorem 3.1}: {\it Let $G$ be an ESL, and $v(k^\mu), v(k'^\mu)\in \mathcal{V}(G)$. In that case:
\begin{equation}
d_G(v(k^\mu),v(k'^\mu)) = max\lbrace\vert k'^\mu - k^\mu\vert\rbrace,
\end{equation}
where $\mu = 0, 1, 2, 3$.}

    Our theorem implies, for instance, that the distance of any vertex of a unit cell from any other vertex is 
exactly one.

    \subsubsection{Eccentricity}

     The {\it eccentricity} $\epsilon_G(v)$ of a vertex $v$ of a graph $G$ is, by definition, the maximum distance
of $v$ from any other vertex. Hence one observes, for example, that the eccentricity of any vertex in a unit cell 
is exactly one.

     Following the ideas taken from the general theory of metric spaces we now define the concept of

     \subsubsection{Ball}

     Let $G$ be a graph, $v$ one of its vertices, and $R$ a positive integer. A {\it ball} $B_G(v,R)$ in $G$ is
defined as a union of all those subgraphs of $G_v$ of $G$, where $v\in \mathcal{V}(G_v)$ and the eccentricity
$\epsilon_{G_v}(v)\le R$. The vertex $v$ is known as the {\it centre} and the positive integer $R$ as the {\it radius}
of the ball $B_G(v,R)$.

    \subsection{Spacetime Lattice (SL)}

    \subsubsection{Isomorphic Graphs}

     The graphs $G$ and $G'$ are called {\it isomorphic}, if there exists a bijective map 
$\psi:\mathcal{V}(G)\longrightarrow\mathcal{V}(G')$ such that for all $u, v \in \mathcal{V}(G)$ $\psi(u)\psi(v)\in E(G')$
if and only if $uv\in E(G)$. In other words, the {\it isomorphism} $\psi$ maps the vertices of $G$ to the vertices of $G'$
in such a way that whenever $u$ and $v$ are vertices of the same edge in $G$, the corresponding vertices $\psi(u)$ and
$\psi(v)$ are vertices of the same edge in $G'$.

     \subsubsection{Connected Graph}

     According to a standard definition, a graph $G$ is {\it connected}, if for arbitrary vertices $u$ and $v$ of
$G$ there exists a path from $u$ to $v$. The concepts of isomorphism and connectedness play an important role in 
the  definition of our spacetime model. Our next step is to define the concept of

     \subsubsection{Spacetime Lattice (SL)}

     A connected graph $G$ is a {\it spacetime lattice (SL)}, if for every vertex $v$ of $G$ there exists a 
positive integer $N_v$, an elementary spacetime lattice $L_v$, and an isomorphism $\psi_v$ from the vertex set
${\mathcal V}(B_G(v,N_v))$ of the ball $B_G(v,N_v)$ to the vertex set ${\mathcal V}(L_v)$ of the elementary 
spacetime lattice $L_v$. 

     In short, our definition of SL states that for every vertex $v$ of a connected graph $G$ there exists a ball
$B_G(v,N_v)$, which is isomorphic to an ESL. The ball $B_G(v,N_v)$ plays a role similar to that of neighbourhood 
in classical spacetime: In classical general relativity one assumes that every point $P$ of spacetime has a 
neighbourhood $U(P)$, which is homeomorphic to $\Re^4$. So we find that classical general relativity is a 
{\it local} theory of space and time in the sense that nothing is assumed about the global properties, such as 
topology, of spacetime. Instead, certain local properties of spacetime are assumed. To preserve this local nature 
of classical general relativity in our model we focus our attention to the balls $B_G(v,N_v)$ surrounding the 
vertices $v$ of the graph $G$ acting as the model of spacetime, instead of considering the graph $G$ itself. For 
each vertex $w$ of the ball $B_G(v,N_v)$ there exists a unique vertex $\psi(w) = v(k^\mu)$ of the corresponding
ESL $L_v$, and vice versa. Hence the coordinates $k^\mu$ of the vertices of $L_v$ may be used as the coordinates
of the corresponding vertices of $B_G(v,N_v)$.

    In classical general relativity spacetime is assumed to be a differentiable manifold and therefore to possess
a differentiable structre. Since in our model spacetime consists of discrete points, and edges joining those points,
it is impossible to introduce the differentiable structure into our model. However, it is possible to introduce a
structure, which plays the role somewhat akin to the differentiable structure of classical spacetimes. With the 
notations adopted in the previous definitions we now introduce the concept of

\subsubsection{Translatory Structure}

  A {\it translatory structure} of a spacetime lattice $G$ is a family of pairs 
$(N_v,\psi_v)_{v\in{\mathcal V}(G)}$ with the following properties:

(i) $G = \bigcup_{v\in{\mathcal V}(G)}B_G(v,N_v)$, 

(ii) For each pair $(B_G(v,N_v),B_G(w,N_w))$, where 
$v, w \in {\mathcal V}(G)$ and\\ $B_G(v,N_v)\cap B_G(w,N_w) \ne \emptyset$ there are four fixed integers
$a^\mu(v,w)$ ($\mu = 0, 1, 2, 3$) such that
\begin{equation}
\psi_w\circ\psi_v^{-1}(v(k^\mu)) = v(k^\mu + a^\mu(v,w))
\end{equation}
for all $v(k^\mu) \in \psi_v({\mathcal V}(B_G(v,N_v)\cap B_G(w,N_w)))$.

\subsubsection{Proper Spacetime Lattice (PSL)}

 A spacetime lattice $G$ is a {\it proper spacetime lattice} (PSL), if it possesses a translatory structure.

 In broad terms our definition of a PSL states that if two balls $B$ and $B'$ in a spacetime lattice intersect 
each other, the coordinates of the vertices of $B'$ may be obtained simply by adding certain constant integers
to their coordinates in $B$. We shall see later that at macroscopic length scales the coordinate $k^0$ plays a 
role of a timelike coordinate, whereas the coordinates $k^{1,2,3}$ are spacelike. In a PSL the coordinates
$k^0$ and $k^{1,2,3}$ preserve their roles as time- and spacelike cordinates, respectively, when we go from
a one ball to another, even if the balls were equipped with different systems of coordinates. 

  One of the reasons for modelling spacetime by a PSL, instead of using an ESL as a model of spacetime, is that 
proper spacetime lattices allow different topologies, whereas the topology of an ESL is trivial. Unless otherwise
stated, however, we shall from this point on consider those subgraphs of a PSL only, which are isomorphic to an 
ESL. In other words, we shall assume that the subgraphs of a PSL under consideration have a trivial topology, and
may be covered by a single system of coordinates.  

    Before being able to define the concept of spacetime itself we must define certain concepts of spacetime 
lattice. Those definitions are given in such subgraph $G'$ of a PSL $G$, which is isomorphic to an ESL. The first 
of these concepts is

      \subsubsection{Line}

     Let $G$ be a PSL, and let $[tk^\mu]$ denote the greatest possible integer, which is smaller than, or equal to
$tk^\mu$ for every $t\in [0, 1]$ and $\mu = 0, 1, 2, 3$. (For example, $[3.8] = 3$.) The 
{\it line} $\lbrace v(k^\mu), v(k'^\mu)\rbrace$ joining the vertices $v(k^\mu)$ and $v(k'^\mu)$ in $G$ is that path
of $G$, which has the vertices
\begin{equation}
v(t) := v([(1-t)k^\mu + tk'^\mu]),
\end{equation}
where $t\in [0, 1]$ such that $max\lbrace\vert k'^\mu-k^\mu\vert\rbrace t$ is an integer.

    To see what this definition means consider, as an example, the line \newline
$\lbrace v(1,3,2,5), v(6,7,5,7)\rbrace$ 
of a PSL. By definition, the vertices of that line are $v(1,3,2,5)$, $v(2,3,2,5)$, $v(3,4,3,5)$, $v(4,5,3,5)$  
$v(5,6,4,6)$ and $v(6,7,5,7)$. 

    According to our definition a line connecting the given two vertices of a SL
(if there is one) is unique. It is easy to see that the following result holds:

    {\bf Theorem 3.2}: {\it The length of a line $\lbrace v(k^\mu), v(k'^\mu)\rbrace$ is 
$max\lbrace\vert k'^\mu - k^\mu\vert\rbrace$.}

     Our theorem implies, for instance, that the line $\lbrace v(1,3,2,5), v(6,7,5,7)\rbrace$ has 5 edges.
Indeed, this is the result which we found above: The line \newline
$\lbrace v(1,3,2,5), v(6,7,5,7)\rbrace$ has 6 vertices,
and therefore it has 5 edges.

     Theorem 3.2 may be used in the proof of Theorem 3.1: It follows from the definition of an ESL that
\begin{equation}
d_G(v(k^\mu), v(k'^\mu)) \ge max\lbrace\vert k'^\mu - k^\mu\vert\rbrace
\end{equation}
for any vertices $v(k^\mu)$ and $v(k'^\mu)$ of an ESL. Since the line $\lbrace v(k^\mu), v(k'^\mu)\rbrace$ is,
according to Theorem 3.2, a path with length $max\lbrace\vert k'^\mu - k^\mu\vert\rbrace$ from a vertex $v(k^\mu)$
to the vertex $v(k'^\mu)$, we find that Eq.(3.6) may be replaced by an exact equality. In other words, we have
proved Theorem 3.1.

   \subsubsection{Linearly Independent Vertices}

    Let $v(k^\mu)$, $v(k'^\mu)$  and $v(k''^\mu)$ be three different vertices of a spacetime lattice $G$. The vertices
$v(k^\mu)$, $v(k'^\mu)$ and $v(k''^\mu)$ are said to be {\it linearly independent} (LI) in $G$, if they do not belong
to the same line in $G$.

   We now define a concept which plays an important role when we reduce the metric properties of spacetime to the
quantum states of the microscopic black holes acting as its constituents:
 
   \subsubsection{Discrete Triangle}

   A {\it discrete triangle} $\lbrace v(k^\mu), v(k'^\mu), v(k''^\mu)\rbrace$ with linearly independent vertices
$v(k^\mu)$, $v(k'^\mu)$ and $v(k''^\mu)$ of a proper spacetime lattice $G$ is a subgraph $G'$ of $G$ such that:

(i) The vertices of $G'$ are of the form:
\begin{equation}
v(s,t) := v([(1-s-t)k^\mu + sk'^\mu + tk''^\mu]),
\end{equation}
where $s \in [0, 1]$ and $t \in [0, 1-s]$ for fixed $s$ such that $max\lbrace\vert k'^\mu - k^\mu\vert\rbrace\,s$
and $max\lbrace\vert k'^\mu - k^\mu\vert\rbrace\,t$ are integers.

(ii) $v(\tilde{k}^\mu)v(\bar{k}^\mu)\in E(G')$ if and only if 
$max\lbrace\vert\tilde{k}^\mu - \bar{k}^\mu\vert\rbrace = 1$.

   The lines $\lbrace v(k^\mu), v(k'^\mu)\rbrace$, $\lbrace v(k'^\mu), v(k''^\mu)\rbrace$ and 
$\lbrace v(k''^\mu), v(k^\mu)\rbrace$ are known as 
the {\it sides} of the discrete triangle $\lbrace v(k^\mu), v(k'^\mu), v(k''^\mu)\rbrace$.

   The simplest possible discrete triangle is the one, where
\begin{equation}
max\lbrace\vert k'^\mu - k^\mu\vert\rbrace = max\lbrace\vert k''^\mu - k^\mu\vert\rbrace 
= max\lbrace\vert k''^\mu - k'^\mu\vert\rbrace = 1
\end{equation}
for all $\mu = 0, 1, 2, 3$. The only vertices of this triangle are $v(k^\mu)$, $v(k'^\mu)$ and $v(k''^\mu)$, and
its sides are edges of the spacetime lattice $G$. In what follows, we shall call this triangle as an {\it 
elementary triangle}. As another example we may consider a triangle 
$\lbrace v(0,0,0,0), v(0,0,3,0), v(0,0,0,3)\rbrace$. Eq.(3.7) implies that the vertices of this triangle are
$v(0,0,0,0)$, $v(0,0,1,0)$, $v(0,0,2,0)$, $v(0,0,3,0)$, $v(0,0,0,1)$, $v(0,0,0,2)$, $v(0,0,0,3)$, $v(0,0,1,1)$,
$v(0,0,2,1)$ and \newline$v(0,0,1,2)$.

   Graph theory is closely related to {\it algebraic topology}. One of the basic concepts of algebraic topology is:

   \subsubsection{Abstract Simplicial Complex (ASC)}

   A set $\mathcal{K}$ of non-empty subsets of a given countable set $\mathcal{W}$ is an 
{\it abstract simplicial complex} (ASC), if \cite{kaksikymmentayksi}

   (i) $\lbrace v\rbrace \in \mathcal{K}$ for all $v\in\mathcal{W}$.

   (ii) If $\sigma\in\mathcal{K}$, then every non-empty subset of $\sigma$ also belongs to $\mathcal{K}$.

  The elements $\sigma$ of $\mathcal{K}$ are known as the {\it simplices} of $\mathcal{K}$, and the elements $v$
of $\mathcal{W}$ as its {\it vertices}. The {\it dimension} $dim(\sigma)$ of a simplex $\sigma$ is one less than
the number of its vertices. If $\tau$ is a subset of $\sigma$ with $k$ vertices $(k\le dim(\sigma) + 1)$, then
$\tau$ is a $(k-1)$-dimensional {\it face} of $\sigma$. The dimension of the abstract simplicial complex 
$\mathcal{K}$ itself is $dim(\mathcal{K}) = sup_{\sigma\in\mathcal{K}}(dim(\sigma))$. A simplex $\tau$ belongs
to the {\it boundary} $\partial\mathcal{K}$ of $\mathcal{K}$, if it is a $k$-dimensional face of exactly one
$(k+1)$-dimensional simplex $\sigma\in\mathcal{K}$ such that $k+1\le dim(\mathcal{K})$.

  It is easy to see that graphs and abstract simplicial complexes are closely related. For instance, if we consider
the edges of a given graph as two-element subsets of its vertex set, we find that the set which consists of the 
vertices and the edges of a given graph is a one-dimensional abstract simplicial complex, which has the vertices 
and the edges of the graph as its simplices. Following this idea we may observe that if we consider an elementary
triangle of a proper spacetime lattice as a three-element subset of the vertex set of that lattice, then a set which 
consists of a collection of elementary triangles, together with their vertices and edges, is a two-dimensional
ASC.

   It turns out useful to define for two-dimensional abstract simplicial complexes the concept of

   \subsubsection{Plate}

   Let $\mathcal{K}$ be a two-dimensional ASC, and $v$ a vertex of $\mathcal{K}$. An ASC 
$\mathcal{P}_v\subset\mathcal{K}$ is a {\it plate}, if $\mathcal{P}_v$ has the following properties:

  (i) Every two-dimensional simplex $\sigma_v\in\mathcal{P}_v$ has $v$ as one of its vertices.

  (ii) If a simplex $\tau\in\mathcal{P}_v$, then $\tau$ is a face of some two-dimensional simplex of 
$\mathcal{P}_v$.

  (iii) If a one-dimensional simplex $\tau_v\in\mathcal{P}_v$ has $v$ as one of its vertices, then $\tau_v$
is a face of exactly two two-dimensional simplices of $\mathcal{P}_v$.
   
   We shall call the vertex $v$ as the {\it centre} of the plate $\mathcal{P}_v$.

   With the concepts of ASC and plate in our service we are able to define a concept, which plays a key role in
our quantum-mechanical model of spacetime:

    \subsubsection{Two-Dimensional Subgraph}

     We shall call a two-dimensional abstract simplicial complex $\Sigma^{(2)}$ a {\it two-dimensional subgraph} 
of a spacetime lattice $G$, if it has the following properties:

   (i) For every two-dimensional simplex $\sigma$ of $\Sigma^{(2)}$ there exists an elementary triangle in $G$ 
such that the vertices of $\sigma$ are the vertices of that elementary triangle.

   (ii) Every one-dimensional simplex, or {\it edge} of $\Sigma^{(2)}$ is a one-dimensional face of either one or
two two-simplices of $\Sigma^{(2)}$. 

   (iii) For every vertex $v$ of $\Sigma^{(2)}$ there exists an edge in $\Sigma^{(2)}$ such that $v$ is a vertex
of that edge. 

   (iv) If $v$ is a vertex of $\Sigma^{(2)}$ such that $\lbrace v\rbrace\not\in\partial\Sigma^{(2)}$, then $v$
is a centre of a plate in $\Sigma^{(2)}$.

   (v) The vertices and the edges of $\Sigma^{(2)}$ constitute a connected subgraph of $G$.

   (vi) The first and the second simplicial homology groups of $\Sigma^{(2)}$ are trivial.

   Our definition of $\Sigma^{(2)}$ has been constructed in such a way that $\Sigma^{(2)}$ corresponds, as closely
as possible, to a smooth, orientable, simply connected two-surface of classical spacetime. Basically, the 
conditions (i) - (v) in our definition state that $\Sigma^{(2)}$ is a surface-like, connected net of elementary 
triangles of $G$. The condition (vi) is a bit more complicated, because it involves simplicial homology groups. The
requirement that the first simplicial homology group of $\Sigma^{(2)}$ must be trivial means that our surface-like
net contains no holes nor handles, whereas the triviality of the second homology group implies that there are no 
attached spheres nor other closed surfaces in $\Sigma^{(2)}$. 

    Our long journey towards a mathematically precise definition of a model of spacetime is now approaching its 
end:

   \subsection{Spacetime}

   In our model {\it spacetime} is defined as a pair $(G,\varphi)$, where $G$ is a proper spacetime lattice, and 
$\varphi: \mathcal{V}(G)\longrightarrow \mathcal{H}_{BH}$ is a map from the vertex set $\mathcal{V}(G)$ of $G$
to the Hilbert space $\mathcal{H}_{BH}$ of the states of a microscopic quantum black hole. In other words, the map
$\varphi$ associates with every vertex $v$ of the spacetime lattice $G$ a unique state 
$\vert\varphi\rangle_v = \varphi(v)$ of a microscopic black hole. If one likes, one may interpret this in such a 
way that at each vertex of a the graph acting as a model of spacetime we have a microscopic quantum black hole
lying on a definite quantum state, which is not necessarily a horizon area eigenstate.  

   \section{Statistics of Spacetime}

   \subsection{Independence Postulates}

   Our model of spacetime introduced in the previous Section gives a rise for several questions: Is it possible to
assign the concept of area with the discrete triangles and the two-dimensional subgraphs of spacetime? How is this
concept of area related to the horizon area eigenstates of the microscopic quantum black holes lying on the 
vertices of the graph acting as the model of spacetime? What is the {\it system} which we should assign with 
specific quantum states in our model? 

   It is easy to see that none of these questions may be answered by means of the properties of our model. Instead, 
those answers must be {\it postulated}. In other words we must introduce, in addition to the general mathematical
definition given in Section 3, certain physical assumptions concerning our model. The first of these assumptions 
are contained in the following {\it independence postulates}:

  (IP1): The quantum states of the microscopic quantum black holes lying on an arbitrary two-dimensional subgraph
$\Sigma^{(2)}$ of spacetime are independent of each other.

  (IP2): The measurement of the area of $\Sigma^{(2)}$ will take the holes on $\Sigma^{(2)}$ to their horizon area
eigenstates.

  (IP3): The ground states of the holes on $\Sigma^{(2)}$ do not contribute to the total area of $\Sigma^{(2)}$.

  (IP4): When a hole is in the $n$'th excited state, it contributes to $\Sigma^{(2)}$ an area, which is directly
proportional to $n$. 

  (IP5): The total area $A$ of $\Sigma^{(2)}$ is the sum of the areas contributed by the black holes on 
$\Sigma^{(2)}$ to the area of $\Sigma^{(2)}$.

   \subsection{Properties of the Independence Postulates}

    An immediate consequence of our independence postulates is that the possible eigenvalues of the area $A$ of a
two-dimensional subgraph $\Sigma^{(2)}$ of are, in SI units, of the form:
\begin{equation}
A = \alpha\ell_{Pl}^2(n_1 + n_2 + ... + n_N).
\end{equation}
In this equation, $\ell_{Pl}$ is the Planck length and the non-negative integers \newline$n_1, n_2,..., n_N$ are the 
quantum numbers associated with the horizon area eigenstates of the $N$ microscopic quantum black holes lying
on the vertices of $\Sigma^{(2)}$. $\alpha$ is a pure number to be determined later. From a purely mathematical 
point of view one may consider Eq.(4.1) as the {\it definition} of the area of $\Sigma^{(2)}$. The physical 
motivation for this definition, however, comes from our independece postulates.

   The first independence postulate (IP1) says that the quantum states of the black holes lying on the
vertices of $\Sigma^{(2)}$ are independent of each other. No doubt this is a rather daring assumption. At this 
stage, however, that assumption is necessary if we want to make progress. Actually, the postulate (IP1) is very
similar to the assumption of asymptotic freedom of quarks in QCD: When quarks come very close to each other, they 
may be considered, essentially, as free and independent particles. Maybe something similar happens for microscopic 
quantum black holes at the Planck scale. As such the postulate (IP1) is not necessarily entirely unphysical. The 
postulate (IP2), in turn, is just an implementation of one of the basic principles of quantum mechanics into our 
model. The principle in question says that only the eigenvalues of observables may be observed. In other words,
measurement collapses the state vector of a system to one of the eigenstates of its observables. As it comes to
the postulate (IP3), that postulate is very similar to one of the basic properties of quantum field theories: In 
ordinary quantum field theories the vacuum states of the particles of the field do not contribute to the total 
energy of the field. Since in our model area takes the place of energy as a quantity labelling the physical states
of the constituents of a system, it is only natural to assume that the vacuum states of the quantum black holes
do not contribute to the area of $\Sigma^{(2)}$. The postulates (IP4) and (IP5) are self-explanating, and 
therefore there is no need to elaborate them any further.

   So far we have talked about two-dimensional subgraphs only. The area associated with those objects is given
by Eq.(4.1). In what follows, we shall assume that Eq.(4.1) holds not only for two-dimensional subgraphs, but also
for discrete triangles as well. In other words, we shall define the area of a discrete triangle with $N$ vertices
to be given by Eq.(4.1).

   \subsection{Statistical Postulates}

   We are now prepared to enter to the statistics of our model. The fundamental problem of the statistical physics
of any system is to identify the micro- and the macrostates of the system. During the past fifteen years 
or so it has become increasingly clear that {\it spacelike two-surfaces} play a fundamental role in gravitational 
physics. One of the first realizations of this possibility was made by 't Hooft in a very influential paper,
where it was suggested that in gravitational physics the observational degrees of freedom may be described
as if they were Boolean variables defined on a two-dimensional lattice evolving in time. \cite{nobelisti} Later,
 it has been shown by Padmanabhan and others that Einstein's field equation may be obtained by varying,
in an appropriate manner, the boundary term of the Einstein-Hilbert action. \cite{kaksikymmentakaksi}  

   Since spacetime in our model is assumed to be a certain graph, the first question is: What is the counterpart 
of a spacelike two-surface in our model? Intuitively, it is fairly obvious that the two-dimensional subgraph
$\Sigma^{(2)}$ introduced in Section 3 might be an appropriate counterpart. The only problem is that we have not introduced
the concept of "time" into our model yet, and therefore it is impossible to say, at least at this stage of 
research, which of the subgraphs $\Sigma^{(2)}$ may be considered "spacelike", and which not. We shall therefore
consider all two-dimensional subgraphs $\Sigma^{(2)}$ of spacetime on an equal footing. For the microscopic 
quantum black holes lying on the vertices of $\Sigma^{(2)}$ we pose the following {\it statistical postulates}:

   (SP1): The microstates of $\Sigma^{(2)}$ are uniquely determined by the combinations of the non-vacuum horizon 
area eigenstates of the microscopic quantum black holes on $\Sigma^{(2)}$.

   (SP2): Each microstate of $\Sigma^{(2)}$ yielding the same area eigenvalue of $\Sigma^{(2)}$ corresponds to the
same macrostate of $\Sigma^{(2)}$.

   The statistics, as well as most parts of the thermodynamics of our model, will be based on these postulates and
 Eq.(4.1).

   \subsection{Statistical Weight and Entropy}

   \subsubsection{Statistical Weight}

    Using our postulates one may calculate the {\it statistical weight} $\Omega$ of a given macrostate of 
$\Sigma^{(2)}$, or the number of microstates of $\Sigma^{(2)}$ corresponding to that macrostate. Eq. (4.1)
implies that the possible area eigenvalues of $\Sigma^{(2)}$ are of the form:
\begin{equation}
A_n = n\alpha\ell_{Pl}^2,
\end{equation}
where the positive integer $n$ is expressed in terms of the non-negative integers $n_1, n_2,..., n_N$ labelling
the horizon area eigenstates of the $N$ microscopic quantum black holes lying on the vertices of $\Sigma^{(2)}$
such that:
\begin{equation}
n = n_1 + n_2 + ... + n_N.
\end{equation}
The postulate (SP2) implies that the macrostates of $\Sigma^{(2)}$ are uniquelly labelled by $n$. In other words,
for each macrostate of $\Sigma^{(2)}$ there exists a unique positive integer $n$, and vice versa. The postulates
(SP1) and (SP2) in together imply that the statistical weight $\Omega(n)$ of the macrostate corresponding to the
positive integer $n$ is the number of ways of writing $n$ as a sum of at most $N$ positive integers. More 
precisely, $\Omega(n)$ is the number of the ordered strings $(n_1,n_2,...,n_m)$, where 
$n_1, n_2,..., n_m\in\lbrace 1, 2, 3,...\rbrace$ and $1\le m\le N$ such that \cite{kaksikymmentakolme}
\begin{equation}
n_1 + n_2 + ... + n_m = n.
\end{equation}

  It is pretty easy to find a general expression for $\Omega(n)$. To begin with, we observe that the number of ways
of writing a given positive integer $n$ as a sum of exactly $m$ positive integers is, when $m\le n$, given by the
binomial coefficient
\begin{equation}
\Omega_m(n) = \left(\begin{array}{cc}n-1\\m-1\end{array}\right).
\end{equation}  
For instance, there are $\left(\begin{array}{cc}4\\2\end{array}\right) = 6$ ways to express a number 5 as a sum of 
exactly 3 positive integers. Indeed, we have:
\begin{equation}
5=3+1+1=1+3+1=1+1+3=1+2+2=2+1+2=2+2+1.
\end{equation}
To see how Eq.(4.5) comes out, consider $n$ identical balls in a row. It is easy to see that $\Omega_m(n)$ is the
number of ways of arranging the $n$ balls in $m$ groups by putting $m-1$ divisions in the $n-1$ empty places between
the balls. There are $\left(\begin{array}{cc}n-1\\m-1\end{array}\right)$ ways of picking up $m-1$ places for the
divisions, and so Eq.(4.5) follows.

  The calculation of $\Omega(n)$ is based on Eq.(4.5). Let us assume that $N$, the number of microscopic quantum
black holes on $\Sigma^{(2)}$, is smaller than $n$. In that case
\begin{equation}
\Omega(n) = \sum_{m=1}^N\Omega_m(n) = \sum_{k=0}^{N-1}\left(\begin{array}{cc}n-1\\k\end{array}\right).
\end{equation}
For instance, there are 
\begin{equation}
\Omega(5) = \left(\begin{array}{cc}4\\0\end{array}\right) + \left(\begin{array}{cc}4\\1\end{array}\right)
            + \left(\begin{array}{cc}4\\2\end{array}\right)
          = 1 + 4 + 6
          = 11
\end{equation}
ways to express the number 5 as a sum of at most 3 positive integers. In the special case, where $N = n$, we have:
\begin{equation}
\Omega(n) = \sum_{k=0}^{n-1}\left(\begin{array}{cc}n-1\\k\end{array}\right) = 2^{n-1}.
\end{equation}
If $N > n$, $\Omega(n)$ is simply the number of ways of writing $n$ as a sum of positive integers, no matter how
many. Since the maximum number of those positive integers is $n$, we find that $\Omega(n)$ is given by Eq.(4.9),
whenever $N \ge n$.

  \subsubsection{Entropy in the Low Temperature Limit}

   As for ordinary thermodynamical systems we define the {\it entropy} of a given macrostate of a two-dimensional
subgraph $\Sigma^{(2)}$ as the natural logarithm of the statistical weight $\Omega(n)$ of that macrostate:
\begin{equation}
S(n) := \ln\Omega(n)
\end{equation}
or, in SI units:
\begin{equation}
S(n) := k_B\ln\Omega(n).
\end{equation}
We first consider the case, where $N>n$. In that case the average value
\begin{equation}
\bar{n} := \frac{n}{N}
\end{equation}
of the quantum numbers $n_1, n_2,..., n_N$ labelling the horizon area eigenstates of the microscopic quantum
black holes on $\Sigma^{(2)}$ is less than one, i. e. the holes are, in average, close to the vacuum state, where
$n_1 = n_2 =...= n_N = 0$. We shall call the case $n < N$ as the {\it low temperature limit} of our model. Of 
course we have not introduced the concept of temperature into our model yet. We shall adopt the name "low 
temperature limit" simply because it is a general feature of physical systems that in low temperatures their 
constituents are close to the ground state.

    Eq.(4.9) implies that in the low temperature limit the entropy of a two-dimensional subgraph $\Sigma^{(2)}$
takes the form:
\begin{equation}
S(n) = (n-1)\ln 2. 
\end{equation}
Using Eq.(4.2) we find that the entropy $S$ may be written in terms of the area $A$ as:
\begin{equation}
S(A) = \frac{\ln 2}{\alpha}A.
\end{equation}
When obtaining Eq.(4.14) we have assumed that the area $A$ is very large, which means that we may replace $n-1$
by $n$ in Eq.(4.13). In the SI units we may write Eq.(4.14) as:
\begin{equation}
S(A) = \frac{\ln 2}{\alpha}\frac{k_Bc^3}{\hbar G}A.
\end{equation}
This is a very satisfying result, because we know from black hole thermodynamics that the entropy of a black hole
horizon is proportional to its area. \cite{nelja, viisi} There are therefore good hopes that our quantum-mechanical  model of spacetime
might be capable to provide a microscopic explanation to this so called Bekenstein-Hawking entropy law. In the next
paper in our series we shall use Eq.(4.15) as one of the starting points of the investigation of the 
thermodynamical properties of spacetime.

   \subsubsection{Entropy in the High Temperature Limit}

   We shall now consider the limit, where $n >> N$, and therefore $\bar{n} >> 1$, which means that the microscopic
black holes on $\Sigma^{(2)}$ are, in average, in highly excited states. Since it is a general feature of all 
physical systems that in high temperatures their constituents are in highly excited states, we shall call the 
limit, where $\bar{n} >> 1$ as the {\it high temperature limit} of our model. When calculating the entropy
of $\Sigma^{(2)}$ in the high temperature limit, we must use Eq.(4.7). Because, in general:
\begin{equation}
\left(\begin{array}{cc}n-1\\k\end{array}\right) = \frac{1}{k!}(n-1)(n-2)...(n-k),
\end{equation}
whenever $k \le n-1$, we find that when $n >> k$, we have:
\begin{equation}
\left(\begin{array}{cc}n-1\\k\end{array}\right) \sim \frac{n^k}{k!}.
\end{equation}
In other words, if the left hand side of Eq.(4.17) is divided by its right hand side, the resulting quotient goes
to unity in the limit, where $n/k\rightarrow\infty$. For instance, for the binomial coefficient 
$\left(\begin{array}{cc}200\\5\end{array}\right)$ the relative error made in the approximation is less than 5 per 
cents. Writing the binomial coefficients on the right hand side of Eq.(4.7) as products, and taking the sum one
finds that the last term in the sum dominates. Hence we may write:
\begin{equation}
\Omega(n) = \sum_{k=0}^{N-1}\left(\begin{array}{cc}n-1\\k\end{array}\right) \sim \frac{1}{(N-1)!}(n-1)^{N-1}.
\end{equation}
This approximation, also, is fairly precise. For instance, when $n = 200$ and $N = 7$, the relative error made in 
the approximation is less than 5 per cents.

   In the limit, where $\bar{n} >> 1$, the entropy of the two-dimensional subgraph $\Sigma^{(2)}$ is:
\begin{equation}
S(n) = \ln\Omega(n) \approx (N - 1)\ln(n-1) - (N-1)\ln(N-1) + N-1,
\end{equation}
where we have assumed that $N$ is very large, and we have used Stirling's formula. The last two terms on the
rigt hand side of Eq.(4.19) are mere additive constants with respect to $n$, and therefore they may be neglected.
($N$ represents here a particle number, and all physically relevant quantities obtainable from the entropy $S$ of
a system are those partial derivatives of $S$, where $N$ has been kept as a constant during the differentiation.)
When we write $S$ as function of $A$, and abandon the physically irrelevant additive constants, we get:
\begin{equation}
S(A) = N\ln A.
\end{equation}
In other words, when the microscopic quantum black holes on the two-dimensional subgraph $\Sigma^{(2)}$ are, in
average, in highly excited states, the entropy is no more proportional to the area $A$, but it depends 
{\it logarithmically} on $A$. It has been speculated for a long time that there might be, in addition to a simple
proportionality, a logarithmic dependence between area and entropy. \cite{kaksikymmentanelja} Our model implies that there indeed exists 
such a dependence, and this logarithmic dependence dominates in a certain limit. It should be noted that when 
obtaining Eq.(4.20) we took into account that $n$ and $N$ are both very large, and therefore we were justified to
replace $n-1$ by $n$ and $N-1$ by $N$. 

   In this paper we consider the entropy $S$ of a two-dimensional subgraph $\Sigma^{(2)}$ in the low- and the high
temperature limits only. A detailed investigation of the general properties of the entropy will be performed, by 
means of the {\it partition function} constructed for $\Sigma^{(2)}$, in the forthcoming papers.

\section{Emergence of Classical from Quantum Spacetime}

   Whatever the correct quantum mechanical model of spacetime may be, it should ultimately
reproduce Einstein's field equation in an appropriate limit. Einstein's field equation,
however, is written in {\it classical} spacetime, and therefore the question is: How does 
classical spacetime emerge from quantum spacetime? Presumably the relationship between 
classical and quantum spacetime is analogous to that of a macroscopic body, and its 
atomic substructure: the properties of a macroscopic body arise as sort of statistical 
averages of the properties of its atoms. In the same way, the properties of classical 
spacetime should arise as statistical averages of its microscopic constituents.

   Classical spacetime is an approximation of quantum spacetime in a long distance and,
presumably, low temperature limit. The fundamental concepts of classical spacetime, such as
metric and curvature, are absent in quantum spacetime. The object of this Section is to find
the precise relationship between the concepts used in classical and quantum spacetimes.

    \subsection{Geometrical Simplices}

    \subsubsection{Basic Definitions}

    In our approach a reduction of the concepts of classical spacetime to the corresponding concepts of
quantum spacetime is achieved by means of the properties of the so called {\it geometrical simplices}. 
\cite{kaksikymmentaviisi}  In 
general, a {\it geometrical n-simplex} $\sigma = v_0v_1...v_n$ is defined as an intersection of all those 
convex sets in $\Re^m$, $(m\ge n)$, which include the $n+1$ linearly independent points $v_0, v_1,..., v_n$
of $\Re^m$. The points $v_0, v_1,..., v_n$ are known as the {\it vertices} of the simplex $\sigma$. A 
zero-simplex $v_0$, for instance, is just a point, a one-simplex $v_0v_1$ a line, a two-simplex $v_0v_1v_2$
a triangle, a three-simplex $v_0v_1v_2v_3$ a solid tetrahedron, and so on. A $k$-simplex $\tau = w_0w_1..w_k$
is known as the $k$-dimensional {\it face} of the simplex $\sigma$, if $\lbrace w_0, w_1,..., w_k\rbrace\subset 
\lbrace v_0, v_1,..., v_n\rbrace$.
 
   As the reader may recall, we talked about {\it abstract simplices} in Section 3. Abstract and geometrical
simplices are closely related in the sense that every $n$-dimensional abstract simplex $\sigma_{abs}$ has 
always a {\it geometrical realization}, or a pair $(f, \sigma_{geom})$, where $\sigma_{geom}$ is a 
geometrical $n$-simplex, and $f$ a bijective map from the vertices of $\sigma_{abs}$ to the vertices of 
$\sigma_{geom}$ such that $f(w_0)f(w_1)...f(w_k)$ is a $k$-dimensional face of $\sigma_{geom}$ if and only
if $\lbrace w_0, w_1,..., w_k\rbrace$ is a $k$-dimensional face of $\sigma_{abs}$ for every $k\le n$.

    \subsubsection{Areas and Edge Lengths}

    Four-dimensional geometrical simplices have a very interesting property: The number of one-faces, or edges, 
of a four-simplex is
\begin{equation}
\left(\begin{array}{cc}5\\2\end{array}\right) = 10,
\end{equation}
which is the same as the number of its two-faces, or triangles:
\begin{equation}
\left(\begin{array}{cc}5\\3\end{array}\right) = 10.
\end{equation}
In other words, there is a one-to-one correspondence between the edges and the triangles of a four-simplex.
This is a phenomenon, which occurs in four dimensions only. It does not happen in any other number of dimensions.
Because of the one-to-one correspondence between the edges and the triangles of a four-simplex it is possible to
write, at least under certain conditions, not only the areas of the triangles in terms of the edge lengths, but
also the edge lengths in terms of the areas.

   The problem is then how to calculate the edge lengths and the triangle areas of an arbitrary geometrical
four-simplex. Recall that a geometrical four-simplex $\sigma = v_0v_1v_2v_3v_4$ is an intersection of all
those convex subsets of $\Re^m$ $(m\ge 4)$, which include the 5 linearly independent points $v_0, v_1, v_2,
v_3, v_4$, and hence the four-simplex $\sigma$ may be understood as a specific four-dimensional convex
subset of $\Re^4$. The lengths of its edges, as well as the areas of its triangles, depend on the metric chosen
in that subset of $\Re^4$. In what follows, we shall assume that the four-simplices under consideration are
equipped with a flat Minkowski metric with a signature $(-,+,+,+)$. We shall show a bit later that in this
case the area of an arbitrary two-face, or triangle $v_av_bv_c$ of a four-simplex $v_0v_1v_2v_3v_4$ 
$(a, b, c \in\lbrace 0, 1, 2, 3, 4\rbrace)$ is
\begin{equation}
A_{abc} = \frac{1}{4}\sqrt{4s_{ac}s_{bc} - (s_{ac} + s_{bc} - s_{ab})^2},
\end{equation}
provided that the triangle $v_av_bv_c$ spacelike, and 
\begin{equation}
A_{abc} = \frac{1}{4}\sqrt{(s_{ac} + s_{bc} - s_{ab})^2 - 4s_{ac}s_{bc}},
\end{equation}
provided that it is timelike. In Eqs.(5.3) and (5.4) $s_{ab}$ denotes the inner product of the vector joining the vertices
$v_a$ and $v_b$ with itself. If the areas $A_{abc}$ are known, Eqs.(5.3) and (5.4) constitute a system of 10 
equations and 10 unknowns $s_{ab}$. Under certain conditions this system of equations has a unique solution, where
$s_{ab}$ is positive, if the edge $v_av_b$ is spacelike, and negative if the edge $v_av_b$ is timelike.

  \subsection{The Correspondence Hypothesis}

  \subsubsection{The Statement of the Hypothesis}

   To complete the assumptions of our quantum-mechanical model of spacetime we shall now express the following
{\it correspondence hypothesis} concerning the areas of discrete triangles in our model:

  Let $v_0 := v(k^\mu)$ be an arbitrary vertex of spacetime, $m$ a positive and $p$ a non-negative integer, 
and let us define the vertex 
\begin{equation}
v_1 := v(k^\mu + (m + p)\delta^\mu_0),
\end{equation}
and the vertices $v_a$ $(a = 2, 3, 4)$ such that
\begin{equation}
v_a := v(k^\mu + m\delta^\mu_{a-1}).
\end{equation}
In that case for every non-negative integer $p$ there exists a positive integer $M_p$ such that for all $m\ge M_p$ the 
abstract four-simplex $\lbrace v_0, v_1, v_2, v_3, v_4\rbrace$ has a geometrical realization 
$\sigma_0 = \tilde{v}_0\tilde{v}_1\tilde{v}_2\tilde{v}_3\tilde{v}_4$ with the following properties:

 (i) $\sigma_0$ is equipped with a flat Minkowski metric.

 (ii) The three-face $\tilde{v}_0\tilde{v}_2\tilde{v}_3\tilde{v}_4$ is spacelike.

 (iii) The edges $\tilde{v}_a\tilde{v}_1$ are timelike or null for every $a = 0, 2, 3, 4$.

 (iv) The area of a triangle $\tilde{v}_a\tilde{v}_b\tilde{v}_c$ of $\sigma_0$ equals to the area of the 
corresponding discrete triangle $\lbrace v_a, v_b, v_c\rbrace$ for every 
$a, b, c \in \lbrace 0, 1, 2, 3, 4\rbrace$. 

\subsubsection{The Meaning of the Hypothesis}

   In very broad terms, our correspondence hypothesis says that at large enough length scales our 
quantum-mechanical model of spacetime will reduce to the classical spacetime. More precisely, at macroscopic length
scales the areas of certain discrete triangles are constrained such that it is possible to construct  
geometrical four-simplices embedded in flat Minkowski spacetime such that the areas of their triangles equal
to the areas of the corresponding discrete triangles. Because the area of a discrete triangle may be expressed
in terms of the quantum numbers $n_1, n_2,..., n_N$ labelling the quantum states of the microscopic quantum 
black holes lying on its $N$ vertices, our hypothesis may be understood as a constraint between the quantum
states of the holes at macroscopic scales. The name "correspondence hypothesis" comes from the fact that our
hypothesis is somewhat analogous to the correspondence principle of quantum mechanics which says, in effect,
that at macroscopic scales quantum physics should reduce to classical physics.

  It is important to notice that because the four-simplex $\sigma_0$ is assumed to be equipped with a flat
Minkowski metric such that some of the edges of $\sigma_0$ are timelike and some spacelike, our hypothesis 
implies an emergence of {\it time} from our model at macroscopic scales. At the Planck scales there are no 
concepts of time and causality whatsoever, but time appears as an entirely statistical quantity in the limit,
where the number of microscopic quantum black holes in the spacetime region under consideration becomes very
large. In this limit the quantum states of the holes are constrained in such a way that the geometrical and
the causal properties of classical spacetime are reproduced.

  One of the disadvantages of our model is that we are compelled to {\it assume} our correspondence hypothesis,
instead of {\it deriving} it from some fundamental quantum-mechanical rules obeyed by the constituents of spacetime
at Planck scales. It remains to be seen, whether the correspondence hypothesis may be obtained as a 
straightforward consequence of a more advanced quantum-mechanical model of spacetime.

   \subsection{The Metric Structure of Spacetime}
  
   \subsubsection{Systems of Coordinates in Classical and Quantum Spacetimes}

    In classical general relativity spacetime is modelled by a continuous set of points, which is reflected by
the fact that the spacetime coordinates $x^\mu$ which are used to identify the points, may take arbitrary real 
values. In contrast, in our quantum-mechanical model the spacetime points are replaced by the vertices of a certain 
graph, and the coordinates $k^\mu$ of those vertices may take only integer values. Since we should able to 
approximate quantum spacetime by a classical spacetime at macroscopic length scales, it is useful to introduce
into our model new coordinates $\chi^\mu$, which are related to the coordinates $k^\mu$ such that:
\begin{equation}
\chi^\mu := \frac{k^\mu}{N^{/4}},
\end{equation}
where $N$ is the number of vertices, or microscopic quantum black holes, in the spacetime region under 
consideration. For macroscopic regions $N$ is presumably of the order $10^{140}$, which means that $N^{1/4}$ is
of the order $10^{35}$. So we see that in macroscopic regions of spacetime the domain of the coordinates 
$\chi^\mu$ becomes, in practice, a continuous set of points and it makes sense to talk, for instance, about
the derivatives of given quantities with respect to the coordinates $\chi^\mu$.

     \subsubsection{Tangent Simplex}

     In classical spacetime one associates with every point of spacetime a certain four-dimensional flat space,
which is kown as the {\it tangent space} of spacetime at that point. One of the properties of a tangent space is
that it approximates spacetime in the neighborhood of a given point at the length scales, where the curvature of
spacetime may be neglected. 

     In our model of spacetime we replace the tangent spaces of classical spacetime by certain  
{\it four-dimensional geometrical simplices} associated with the vertices of spacetime. For an arbitrary
vertex $v_0 := v(k^\mu)$ the four-dimensional geometrical simplex in question is the simplex $\sigma_0$ defined 
in our correspondence hypothesis. For the sake of brevity and simplicity we shall henceforth call the four-simplex
$\sigma_0$ associated with the vertex $v_0$ as the {\it tangent simplex} of the vertex $v_0$. According to our
correspondence hypothesis every vertex of spacetime has a tangent simplex. Since the number of edges of a tangent
simplex equals to the number of its triangles, the lengths of its edges may be expressed in terms of the areas of
its triangles and, moreover, in terms of the quantum numbers $n_1, n_2,...$ labelling the quantum states 
of the black holes lying on the vertices of the discrete triangles corresponding to the triangles of the tangent
simplex. In other words, the edge lengths of the tangent simplices may be reduced to the quantum states of certain
quantum black holes.

    \subsubsection{The Metric}

    Consider now a region of spacetime with $N$ vertices, and let us fix the positive integer $m$ of Eqs.(5.5) 
and (5.6) at each vertex of this region such that
\begin{equation}
m = N^{1/4}.
\end{equation}
Assuming that $N$ is large enough our correspondence hypothesis implies that each vertex of the region under
consideration has a tangent simplex. Moreover, if we pick up an arbitrary vertex $v_0 = v(k^\mu)$ with coordinates
$\chi^\mu(v_0) = \frac{k^\mu}{N^{1/4}}$ from that region, then the coodinates of the vertices $v_1$, $v_2$, $v_3$ 
and $v_4$ defined in Eqs.(5.5) and (5.6) are:
\begin{subequations}
\begin{eqnarray}
\chi^\mu(v_1) &=& \chi^\mu(v_0) + (1 + \frac{p}{N^{1/4}})\delta^\mu_0,\\
\chi^\mu(v_a) &=& \chi^\mu(v_0) + \delta^\mu_{a-1}
\end{eqnarray}
\end{subequations}
for all $a = 2, 3, 4$. If we assume that $p$ is very much smaller than $N^{1/4}$, we may write, in effect:
\begin{equation}
\chi^\mu(v_a) = \chi^\mu(v_0) + \delta^\mu_{a-1}
\end{equation}
for all $a = 1, 2, 3, 4$.

  As the next step we define the coordinates $\tilde{\chi}^\mu$ $(\mu = 0, 1, 2, 3)$ on the tangent simplex associated with
the vertices of our region. Let $v_0 = v(k^\mu)$ be an arbitrary vertex and 
$\sigma_0 = \tilde{v}_0\tilde{v}_1\tilde{v}_2\tilde{v}_3\tilde{v}_4$ a tangent simplex associated with that
vertex. We define the coordinates $\tilde{\chi}^\mu(P)$ of an arbitrary point $P$ of $\sigma_0$ such that the position
vector of the point $P$ with respect to $\tilde{v}_0$ is: 
\begin{equation}
\vec{r}(P) = (\tilde{\chi}^\mu(P) - \chi^\mu(v_0))\vec{b}_\mu,
\end{equation}
where we have used Einstein's sum rule, and $\vec{b}_\mu$ is a vector from the vertex $\tilde{v}_0$ to the 
vertex $\tilde{v}_{\mu+1}$ for all $\mu = 0, 1, 2, 3$. Eq.(5.11) implies that the coordinate curves corresponding
to the coordinates $\tilde{\chi}^\mu$ in $\sigma_0$ are straight lines parallel to those edges in $\sigma_0$, which have the
vertex $\tilde{v}_0$ in common. Moreover, Eqs.(5.10) and (5.11) imply that the coordinates of a vertex 
$\tilde{v}_a$ of $\sigma_0$ are
\begin{equation}
\tilde{\chi}^\mu(\tilde{v}_a) = \chi^\mu(v_a)
\end{equation}
for all $a = 0, 1, 2, 3, 4$. In other words, the coordinates $\tilde{\chi}^\mu(\tilde{v}_a)$ of the vertices $\tilde{v}_a$ 
of the tangent simplex $\sigma_0$ coincide with the coordinates $\chi^\mu(v_a)$ of the corresponding vertices of
our spacetime model. This is something to be desired, if we want to approximate a macroscopic region of spacetime
by a tangent simplex.

  Eq.(5.11) now implies that the tangent vectors of the coordinate curves corresponding to the coordinates 
$\tilde{\chi}^\mu$ are:
\begin{equation}
\frac{\partial\vec{r}}{\partial \tilde{\chi}^\mu} = \vec{b}_\mu
\end{equation}
for all $\mu = 0, 1, 2, 3$, and therefore we may write the components of the metric tensor in the coordinate base
determined by the coordinates $\tilde{\chi}^\mu$ as:
\begin{equation}
g_{\mu\nu} = \vec{b}_\mu\bullet\vec{b}_\nu,
\end{equation}
where the dot denotes the inner product between the vectors $\vec{b}_\mu$ and $\vec{b}_\nu$. Since the spacetime
inside the tangent simplex $\sigma_0$ is assumed to be flat and Minkowskian, the inner product is determined by the
flat Minkowski metric with the signature $(-,+,+,+)$. Because the vector from the vertex $\tilde{v}_{\mu+1}$ to
the vertex $\tilde{v}_{\nu+1}$ is
\begin{equation}
\vec{b}_{\mu\nu} := \vec{b}_\nu - \vec{b}_\mu,
\end{equation}
we find, using the standard properties of the inner product:
\begin{equation}
\vec{b}_{\mu\nu}\bullet\vec{b}_{\mu\nu} = \vec{b}_\mu\bullet\vec{b}_\mu + \vec{b}_\nu\bullet\vec{b}_\nu - 
2\vec{b}_\mu\bullet\vec{b}_\nu.
\end{equation}
Hence we observe that if we denote
\begin{subequations}
\begin{eqnarray}
S_\mu :&=& \vec{b}_\mu\bullet\vec{b}_\mu,\\
S_{\mu\nu} :&=& \vec{b}_{\mu\nu}\bullet\vec{b}_{\mu\nu},
\end{eqnarray}
\end{subequations}
the components of the metric tensor at the tangent simplex associated with the vertex with coordinates 
$\chi^\alpha$ are:
\begin{equation}
g_{\mu\nu}(\chi^\alpha) = \frac{1}{2}(S_\mu(\chi^\alpha) + S_\nu(\chi^\alpha) - S_{\mu\nu}(\chi^\alpha))
\end{equation}
for all $\mu, \nu = 0, 1, 2, 3$. If we denote by $s_{ab}$ the inner product of a vector joining a vertex 
$\tilde{v}_a$ to the vertex $\tilde{v}_b$ with itself for every $a, b = 0, 1, 2, 3, 4$, we find that the 
relationship between the quantities $S_\mu$, $S_{\mu\nu}$ and $s_{ab}$ is (Einstein's sum rule has been used):
\begin{subequations}
\begin{eqnarray}
S_\mu &=& s_{0a}\delta^a_{\mu +1},\\
S_{\mu\nu} &=& s_{ab}\delta^a_{\mu+1}\delta^b_{\nu+1}
\end{eqnarray}
\end{subequations}
for all $\mu, \nu = 0, 1, 2, 3$. Using Eq.(5.19) we may write the components $g_{\mu\nu}$ of the metric tensor 
by means of the quantities $s_{ab}$, which are related to the edge lengths of the tangent simplex $\sigma_0$. The
relationship between the independent components of $g_{\mu\nu}$ and the quantities $s_{ab}$ may be written in a 
matrix form:
\begin{equation}
\left(\begin{array}{cc}g_{00}\\g_{01}\\g_{02}\\g_{03}\\g_{11}\\g_{12}\\g_{13}\\g_{22}\\g_{23}\\g_{33}
\end{array}\right)
= \frac{1}{2}\left(\begin{array}{cccccccccc}
                                2&0&0&0&0&0&0&0&0&0\\
                                1&1&0&0&-1&0&0&0&0&0\\
                                1&0&1&0&0&-1&0&0&0&0\\
                                1&0&0&1&0&0&0&-1&0&0\\
                                0&2&0&0&0&0&0&0&0&0\\
                                0&1&1&0&0&0&0&-1&0&0\\
                                0&1&0&1&0&0&0&0&-1&0\\
                                0&0&2&0&0&0&0&0&0&0\\
                                0&0&1&1&0&0&0&0&0&-1\\
                                0&0&0&2&0&0&0&0&0&0
\end{array}\right)
\left(\begin{array}{cc}s_{01}\\s_{02}\\s_{03}\\s_{04}\\s_{12}\\s_{13}\\s_{14}\\s_{23}\\s_{24}\\s_{34}
\end{array}\right),
\end{equation}
or, in a more compact expression:
\begin{equation}
g_{\mu\nu} = B_{\mu\nu}^{ab}s_{ab},
\end{equation}
where $B_{\mu\nu}^{ab}$ denotes the elements of the matrix, which maps the quantities $s_{ab}$ to the independent 
components of $g_{\mu\nu}$, and the repeated indices are summed over.

  Now, using the fact that the area of an arbitrary two-surface may be written, in general, as:
\begin{equation}
A = \int\sqrt{\vert g\vert}\,d^2x,
\end{equation}
where $g$ is the determinant of the metric induced on the surface, and the integration is performed over the 
surface, we find that the area of a triangle $\tilde{v}_a\tilde{v}_b\tilde{v}_c$ of the tangent simplex
$\sigma_0$ is given by Eqs.(5.3) and (5.4) for all $a, b = 0, 1, 2, 3, 4$. Since Eqs.(5.3) and (5.4) constitute a
system of 10 equations and 10 unknowns $s_{ab}$, we may solve the quantities $s_{ab}$ in terms of the areas, and
substitute these solutions in Eq.(5.21) to get the components of the metric tensor. So we have shown how the 
components of the metric tensor may be reduced to the areas of the triangles of the tangent simplex $\sigma_0$,
and hence to the quantum states of the microscopic quantum black holes constituting spacetime. In other words, we
have shown how the metric properties of spacetime at macroscopic scales may be reduced to the quantum-mechanical 
properties of its constituents at microscopic scales. Since the metric tensor is the fundamental object of 
classical general relativity, we have shown how classical spacetime emerges from quantum spacetime at large length
scales.

   Keeping in mind that $g_{\mu\nu}$ is a concept which is meaningful at macroscopic scales only, we may define, by
means of $g_{\mu\nu}$, other quantities familiar from classical general relativity.  For instance, one may define 
the {\it Christoffel symbol}
\begin{equation}
\Gamma^\alpha_{\mu\nu} := \frac{1}{2}g^{\alpha\sigma}
(\frac{\partial g_{\sigma\mu}}{\partial\chi^\nu} 
+ \frac{\partial g_{\nu\sigma}}{\partial\chi^\mu} 
- \frac{\partial g_{\mu\nu}}{\partial\chi^\sigma}).
\end{equation}
When the number of vertices under consideration is enormous, the allowed values of $\chi^\mu$ constitute, in
practice, a continuous set of points, and the differences may be replaced, as an excellent
approximation, by differentials. Hence Eq.(5.23) really makes sense at macroscopic length
scales. For the same reason we may meaningfully talk about diffeomorphism symmetry at 
macroscopic length scales. By means of $\Gamma^\alpha_{\mu\nu}$ and $g_{\mu\nu}$ one may define, in the usual 
way, the Riemann and the Ricci tensors, curvature scalars, covariant derivatives, and so 
on. However, it should always be kept in mind that all these objects are meaninful at macroscopic scales only.           

\subsubsection{A Note on Scale}

   Our construction of the metric of spacetime was based on the notion of tangent simplex: The correspondence 
hypothesis implied that if the positive integer $m$ in Eqs.(5.5) and (5.6) is large enough for an arbitrary 
vertex $v_0$, the vertex $v_0$ may be associated with a tangent simplex such that $v_0$ is one of the vertices
of that tangent simplex, and $m$ gives the graph theoretic distances of the other vertices from $v_0$. The metric
of spacetime at the vertex $v_0$ was determined by the areas of the two-faces of the tangent simplex and those 
areas, in turn, were determined by the quantum states of certain microscopic black holes. 

   It is important to note that the tangent simplex associated with the vertex $v_0$ is determined by the integer 
$m$. If $m$ is changed, so is the tangent simplex, and the components of the metric tensor calculated by means of 
the two-face areas of the "new" tangent simplex are not necessarily the same as were the components calculated
by means of the two-face areas of the "old" tangent simplex. In other words, the metric tensor of spacetime depends
on the {\it scale} determined by the positive integer $m$. This means that in our model it makes no sense to say
that the metric of spacetime at a given vertex $v_0$ is this or that unless we, at the same time, specify the 
integer $m$, and hence the scale, at which we investigate the metric properties of spacetime. For instance, it is
possible that when $m$ is still fairly small but yet large enough such that we may construct a metric for 
spacetime, the foam-like properties may play a dominant role, and there are huge fluctuations in the metric, when 
we go from a one point of spacetime to another, even when those points are relatively close to each other. 
Most likely, these fluctuations will smoothen out, when we
increase $m$, and when $m$ is of the order $10^{35}$, which means that we consider spacetime at one meter scales,
spacetime appears more or less flat. Finally, when $m$ is very much larger than $10^{35}$, the large scale 
curvature of spacetime begins to show up in the metric.

\subsection{The Linear Field Approximation}

  The ideas presented in this Section may be made more concrete, if we consider the {\it linear field
approximation} of classical general relativity. In this approximation the spacetime metric is written in the
form:
\begin{equation}
g_{\mu\nu} = \eta_{\mu\nu} + h_{\mu\nu},
\end{equation}
where $\eta_{\mu\nu} := diag(-1,1,1,1)$ is the flat Minkowski metric in flat Minkowski coordinates. $h_{\mu\nu}$
may be understood as a small perturbation in the flat spacetime metric. When considering the linear field 
approximation in the context of our model the key problem is to find an expression for $h_{\mu\nu}$ in terms
of the quantum states of the microscopic black holes constituting spacetime.

   Our starting point is an observation that the numbers of the vertices on the discrete triangles 
$\lbrace v_a,v_b,v_c\rbrace$ associated with the given vertex $v_0 = v(k^\mu)$ are equal for all $a, b, c \in
\lbrace 0, 1, 2, 3, 4\rbrace$. More precisely, if $p = 0$ in Eq.(5.5), then for fixed $m$ in Eqs.(5.5) and (5.6)
the number of vertices on an arbitrary discrete triangle is
\begin{equation}
\mathcal{N} = \frac{(m+1)(m+2)}{2}.
\end{equation}
To see how this result comes out consider, for example, the discrete triangle $\lbrace v_0,v_1,v_2\rbrace$. Our
definition of the concept of discrete triangle implies that the length of the line 
$\lbrace v(k^0 + q,k^1,k^2,k^3), v(k^0,k^1 + q,k^2,k^3)\rbrace$ is $q$ for all $q = 0, 1, 2,..., m$, 
and therefore it has $q + 1$ vertices. The lines associated with different values
of $q$ do not intersect each other, and Eq.(3.7) implies that those lines include all vertices of the 
discrete triangle $\lbrace v_0,v_1,v_2\rbrace$, when $q$ goes through all integer values from 0 to $m$. Hence we 
find that the number of vertices of the discrete triangle $\lbrace v_0,v_1,v_2\rbrace$ is
\begin{equation}
\mathcal{N} = 1 + 2 + 3 + ... + (m + 1),
\end{equation}
which readily implies Eq.(5.25). A similar reasoning holds for all discrete triangles $\lbrace v_a,v_b,v_c\rbrace$.

   Since the numbers of the vertices on the discrete triangles are equal, one might expect that when the spacetime
region under consideration is in thermal equilibrium, the areas of those triangles, and hence the two-face areas
of the corresponding tangent simplex $\sigma_0$, should be more or less equal. The possible differences in the 
areas are caused by the quantum fluctuations in the horizon area eigenvalues of the microscopic black holes
constituting those discrete triangles. When expressing the perturbation $h_{\mu\nu}$ in terms of the quantum
states of microscopic black holes, we should therefore begin with a tangent simplex with equal two-face areas. The
quantum fluctuations in the horizon area eigenstates will cause small perturbations in the areas of its two-faces,
or triangles, and therefore in its edge lengths. The perturbations in the edge lengths, in turn, will cause 
perturbations in the metric inside the tangent simplex, and hence we shall be able to reduce the small perturbation
$h_{\mu\nu}$ in the flat spacetime metric to the small fluctuations in the quantum states of the constituents of
spacetime.

    Using Eqs.(5.3) and (5.4) we find that if we take
\begin{equation}
s_{a1} = -\frac{1}{2}L^2
\end{equation}
for all $a = 0, 2, 3, 4$ and
\begin{equation}
s_{ab} = L^2,
\end{equation}
whenever $a, b \neq 1$, the area of an arbitrary two-face $\tilde{v}_a\tilde{v}_b\tilde{v}_c$ of a geometrical
four-simplex $\sigma_0 = \tilde{v}_0\tilde{v}_1\tilde{v}_2\tilde{v}_3\tilde{v}_4$ acting as a tangent simplex of
spacetime is:
\begin{equation}
A_{abc} = \frac{\sqrt{3}}{4}L^2,
\end{equation}
no matter, whether the triangle $\tilde{v}_a\tilde{v}_b\tilde{v}_c$ is space- or timelike. In other words,
the two-face areas of the tangent simplex $\sigma_0$ are equal. In Eqs.(5.27) and (5.28) $L$ is an arbitrary
(sufficiently large) positive real number. It has been shown in Appendix A that small changes $\delta A_{abc}$
in the two-face areas $A_{abc}$ of the tangent simplex $\sigma_0$ will cause certain small changes $\delta s_{ab}$
in its squared edge lengths $s_{ab}$ such that the relationship between the small changes $\delta A_{abc}$ and
$\delta s_{ab}$ may be written in the following matrix form:
\begin{equation}
\left(\begin{array}{cc}\delta s_{01}\\\delta s_{02}\\\delta s_{03}\\\delta s_{04}\\\delta s_{12}\\\delta s_{13}\\
\delta s_{14}\\\delta s_{23}\\\delta s_{24}\\\delta s_{34}\end{array}\right)
= \frac{\sqrt{3}}{3}
\left(\begin{array}{cccccccccc}
                       -4&-4&-4&4&4&4&2&2&2&-8\\
                        2&-1&-1&4&4&-2&-1&-1&2&-2\\
                       -1&2&4&4&-2&4&-1&2&-1&-2\\
                       -1&-1&2&-2&4&4&2&-1&-1&-2\\
                       -4&2&2&4&4&-8&-4&-4&2&4\\
                        2&-4&2&4&-8&4&-4&-4&-4&4\\
                        2&2&-4&-8&4&4&2&-4&-4&4\\
                       -1&-1&2&4&-2&-2&2&-1&-1&4\\
                       -1&2&-1&-2&4&-2&-1&2&-1&4\\
                        2&-1&-1&-2&-2&4&-1&-1&2&4   
\end{array}\right)
\left(\begin{array}{cc}\delta A_{012}\\\delta A_{013}\\\delta A_{014}\\\delta A_{023}\\\delta A_{024}\\
\delta A_{034}\\\delta A_{123}\\\delta A_{124}\\\delta A_{134}\\\delta A_{234}\end{array}\right).
\end{equation}
In what follows, we shall write Eq.(5.30) in a more condensed form:
\begin{equation}
\delta s_{ab} = M^{cde}_{ab}\delta A_{cde}.
\end{equation}
where $M^{cde}_{ab}$ denotes the elements of the matrix, which multiplies a column matrix with the elements 
$\delta A_{abc}$, and the repeated indices up and down are summed over.

   If one replaces the quantities $s_{ab}$ in Eq.(5.21) by the quantities $s_{ab} + \delta s_{ab}$, where the
quantities $s_{ab}$ are given by Eqs.(5.27) and (5.28), and the quantities $\delta s_{ab}$ by Eq.(5.31) one
finds, by means of Eqs.(5.20) and (5.21), the components of the metric tensor $g_{\mu\nu}$ in terms of the
quantities $\delta A_{abc}$. However, there is a very grave disadvantage with the metric $g_{\mu\nu}$ obtained
by means of this process: $g_{\mu\nu}$ has been written in a system of coordinates, where it does not reduce to
the flat Minkowski metric $\eta_{\mu\nu}$ in the special case, where
\begin{equation}
\delta A_{abc} \equiv 0
\end{equation}
for all $a, b, c \in \lbrace 0, 1, 2, 3, 4\rbrace$ at every vertex of spacetime. We should therefore replace the
coordinates $\chi^\mu$ by the new coordinates $x^\mu$ such that
\begin{equation}
\frac{\partial \chi^\alpha}{\partial x^\mu}\frac{\partial \chi^\beta}{\partial x^\nu}g_{\alpha\beta} 
\equiv \eta_{\mu\nu},
\end{equation}
whenever Eq.(5.32) holds. It turns out that an appropriate system of coordinates is the one, where the old
coordinates $\chi^\mu$ are given in terms of the new coordinates $x^\mu$ such that:
\begin{subequations}
\begin{eqnarray}
\chi^0 &=& \frac{2\sqrt{14}}{7L}x^0,\\
\chi^1 &=& -\frac{\sqrt{14}}{14L}x^0 + \frac{1}{L}x^1 - \frac{\sqrt{3}}{3L}x^2 - \frac{\sqrt{6}}{6L}x^3,\\
\chi^2 &=& -\frac{\sqrt{14}}{14L}x^0 + \frac{2\sqrt{3}}{3L}x^2 - \frac{\sqrt{6}}{6L}x^3,\\
\chi^3 &=& -\frac{\sqrt{14}}{14L}x^0 + \frac{\sqrt{6}}{2L}x^3.
\end{eqnarray}
\end{subequations}

   Using Eqs. (5.20), (5.21), (5.24), (5.30) and (5.31) one finds that the components of the perturbation 
$h_{\mu\nu}$ in the flat Minkowski metric $\eta_{\mu\nu}$ are:
\begin{equation}
h_{\mu\nu}(x^\alpha) = N^\rho_\mu N^\sigma_\nu B^{ab}_{\rho\sigma}M^{cde}_{ab}\frac{\delta A_{cde}(x^\alpha)}{A},
\end{equation}
where we have defined:
\begin{equation}
N^\rho_\mu := \frac{3^{1/4}}{2}L\frac{\partial \chi^\rho}{\partial x^\mu},
\end{equation}
$\delta A_{abc}(x^\alpha)$  denotes the area change of the discrete triangle $\lbrace v_a, v_b, v_c\rbrace$
associated with the vertex $v_0$ identified by means of the coordinates $x^\alpha$, and $B^{ab}_{\rho\sigma}$
was defined in Eqs.(5.20) and (5.21). Since the area $A_{abc}$ of a
given discrete triangle $\lbrace v_a, v_b, v_c\rbrace$ is, according to Eq.(4.1), proportional to the sum $n_{abc}$
of the quantum numbers labelling the horizon area eigenstates of the microscopic black holes lying on the
vertices of $\lbrace v_a, v_b, v_c\rbrace$, Eq.(5.35) may be written as:
\begin{equation}
h_{\mu\nu}(x^\alpha) = N^\rho_\mu N^\sigma_\nu B^{ab}_{\rho\sigma}M^{cde}_{ab}\frac{\delta n_{cde}(x^\alpha)}{n},
\end{equation}
where
\begin{equation}
n := \frac{A}{\alpha\ell_{Pl}^2}
\end{equation}
is the original sum of those quantum numbers, and $\delta n_{abc}(x^\alpha)$ denotes the change in this sum at the discrete 
triangle $\lbrace v_a, v_b, v_c\rbrace$. So we have found in which way the metric of spacetime will depend, in the
linear field approximation, on the quantum states of its constituents. Explicit expressions for the components of
$h_{\mu\nu}$ in terms of $\delta n_{abc}$ and $n$ are pretty involved, and not very illuminating. Therefore 
we shall not present those explicit expressions in here. An interested reader may calculate them 
straightforwardly by means of Eq.(5.37).

\section{Concluding Remarks}

  In this paper we have considered a specific microscopic model of spacetime, where Planck 
size quantum black holes were used as the fundamental constituents of space and time. 
Spacetime was assumed to a graph, where black holes lie on the vertices. The only physical
degree of freedom associated with a microscopic quantum black hole acting as a fundamental 
building block of spacetime was taken to be its horizon area, which was assumed to have 
a discrete spectrum with an equal spacing. Our idea was to reduce all properties of 
spacetime back to the event horizon area eigenvalues of Planck size quantum black holes.

   We focussed our attention at certain specific subgraphs of spacetime, which we called
{\it two-dimensional subgraphs}, and which may be viewed, at least to some extent, as graph theoretic 
analogues of two-surfaces of classical spacetime. For the Planck size quantum 
black holes lying on the vertices of a two-dimensional subgraph we introduced four independence-,
and two statistical postulates. These postulates implied, among other things, that the
area eigenvalues of a two-dimesional subgraph are of the form:
\begin{equation}
A_n = \alpha n\ell_{Pl}^2,
\end{equation}
where $n=0,1,2,3,...$, $\ell_{Pl}$ is the Planck length, and $\alpha$ is a pure number of
order unity. We also found that two-dimensional subgraphs possess entropy which, in the low 
temperature limit where most black holes on the spacelike two-graph are in their ground
states takes, in the SI units, the form:
\begin{equation}
S = \frac{\ln 2}{\alpha}\frac{k_Bc^3}{\hbar G}A,
\end{equation}
where $A$ is the area of the subgraph. In other words, we found that the entropy of a 
two-dimesional subgraph is proportional to its area.

   In the long distance limit we showed how classical spacetime as such as we know it from 
classical general relativity emerges from our microscopic model of spacetime, and how the 
metric properties of classical spacetime may be reduced to the horizon area eigenstates of 
Planck size black holes. 

  One of the properties our model is that it provides a new interpretation for the concepts
of time and causality. In our model the causal properties of spacetime are determined by 
the statistical distribution of the quantum states of the Planck size black holes constituting
spacetime. More precisely, at macroscopic scales the horizon area eigenstates of the black
holes are distributed in such a way that the metric of the resulting spacetime has a signature $(-,+,+,+)$.
At the Planck scales there are no notions of time nor causality, but time appears as an entirely
statistical concept in the limit, where the number of black holes in the region of spacetime under
consideration is very large. It would be very interesting to investigate, whether our statistical 
concept of time could be related in one way or another to the second law or thermodynamics.

  Although our model of spacetime may meet with some success in the sense that it reduces to classical 
spacetime at macroscopic scales, and gives for the entropy of two-surfaces an expression, which is
consistent with the Bekenstein-Hawking entropy law, it also has problems of its own. The most serious 
of those problems is that we were forced to {\it assume} the so called correspondence hypothesis,
which was one of the key elements of our model, instead of deriving that hypothesis from the fundamental
properties of the constituents of spacetime. In very broad terms, our correspondence hypothesis
stated that at sufficiently large scales the quantum states of the black holes constituting spacetime 
are distributed in such a way that the spacetime region under consideration may be approximated by
a flat four-simplex equipped with a flat Minkowski metric. It remains to be seen, whether that hypothesis
may be obtained as a natural consequence of a more advanced quantum-mechanical model of spacetime.
 
  The main lesson one may learn from this paper is that it really seems to be possible to
construct a mathematically well-defined model of spacetime which, at the Planck 
length scales, takes into account the quantum mechanical properties of spacetime, and reduces 
to the classical spacetime at the large length scales. The question is, however, whether
the model will really be able to reproduce all of the "hard facts" of gravitational physics 
as such as we know them today. These hard facts include, among other things, the Hawking and
the Unruh effects, together with Einstein's field equation. All these things, however, are
more or less connected with the {\it thermodynamics} of spacetime. Indeed, one of the 
starting points of this paper was Jacobson's discovery of the year 1995 that Einstein's
field equation may be understood as a thermodynamical equation of state of spacetime and
matter fields.\cite{yksi} So we see that to recover the hard facts of gravitational physics from our
model we should abandon for a while the quantum mechanical, and even the statistical 
properties of spacetime, and to consider its thermodynamical properties, instead. This will 
be the subject of the next paper in our series.

\appendix

\section{The Derivation of Eq.(5.30)}

 Eqs.(5.3) and (5.4) imply that the partial derivatives of the area $A_{abc}$ of a two-face 
$\tilde{v}_a\tilde{v}_b\tilde{v}_c$ of a tangent simplex $\sigma_0$ associated with the vertex $v_0 = v(k^\mu)$
of spacetime with respect to the squared edge length $s_{ab}$ is
\begin{equation}
\frac{\partial A_{abc}}{\partial s_{ab}} = \frac{s_{ac} + s_{bc} - s_{ab}}{16A_{abc}},
\end{equation}
if the two-face $\tilde{v}_a\tilde{v}_b\tilde{v}_c$ is spacelike, and 
\begin{equation}
\frac{\partial A_{abc}}{\partial s_{ab}} = -\frac{s_{ac} + s_{bc} - s_{ab}}{16A_{abc}},
\end{equation}
if it is timelike. In general, a small change $\delta A_{abc}$ in the area of a given two-face may be written
in terms of the small changes $\delta s_{ab}$ in its squared edge lengths as:
\begin{equation}
\delta A_{abc} = \frac{1}{2}\frac{\partial A_{abc}}{\partial s_{de}}\,\delta s_{de},
\end{equation}
where the repeated indices are summed over. The factor $1/2$ must be included, because $s_{de}$ is symmetric. In 
the special case, where Eqs.(5.26) and (5.27) hold, Eqs.(5.28), (A.1), (A.2) and (A.3) imply:
\begin{equation}
\delta A_{abc} = \frac{\sqrt{3}}{12}(\delta s_{ac} + \delta s_{bc} + \delta s_{ab}),
\end{equation}
if $a, b, c \ne 1$, and 
\begin{equation}
\delta A_{1ab} = \frac{\sqrt{3}}{12}(-\delta s_{1a} + 2\delta s_{ab} - \delta s_{1b}).
\end{equation}
Using Eqs.(A.4) and (A.5) we find that the relationship between the quantities $\delta A_{abc}$ and $\delta s_{ab}$
may be written in a matrix form:
\begin{equation}
\left(\begin{array}{cc}\delta A_{012}\\\delta A_{013}\\\delta A_{014}\\\delta A_{023}\\\delta A_{024}\\
\delta A_{034}\\\delta A_{123}\\\delta A_{124}\\\delta s_{134}\\\delta A_{234}\end{array}\right)
= \frac{\sqrt{3}}{12} 
\left(\begin{array}{cccccccccc}
                       -1&2&0&0&-1&0&0&0&0&0\\
                       -1&0&2&0&0&-1&0&0&0&0\\
                       -1&0&0&2&0&0&-1&0&0&0\\
                        0&1&1&0&0&0&0&1&0&0\\
                        0&1&0&1&0&0&0&0&1&0\\
                        0&0&1&1&0&0&0&0&0&1\\
                        0&0&0&0&-1&-1&0&2&0&0\\
                        0&0&0&0&-1&0&-1&0&2&0\\
                        0&0&0&0&0&-1&-1&0&0&2\\
                        0&0&0&0&0&0&0&1&1&1   
\end{array}\right)
\left(\begin{array}{cc}\delta s_{01}\\\delta s_{02}\\\delta s_{03}\\\delta s_{04}\\\delta s_{12}\\
\delta s_{13}\\\delta s_{14}\\\delta s_{23}\\\delta s_{24}\\\delta s_{34}\end{array}\right).
\end{equation}
When this relationship is inverted, we get Eq.(5.30).

\maketitle

\end{document}